\documentclass[10pt,twocolumn,letterpaper]{article}

\usepackage[pagenumbers]{cvpr} 
\usepackage{caption}
\usepackage{graphicx}
\usepackage{amsmath}
\usepackage{amssymb}
\usepackage{booktabs}
\usepackage{bm}
\usepackage{multirow}
\usepackage{xcolor}         

\usepackage{ulem}
\usepackage{booktabs} 
\usepackage{adjustbox}
\usepackage[utf8]{inputenc} 
\usepackage[T1]{fontenc}    
\usepackage{url}            
\usepackage{booktabs}       
\usepackage{amsfonts}       
\usepackage{nicefrac}       
\usepackage{microtype}      
\usepackage{xcolor}         
\usepackage{xspace}
\usepackage{amsmath}
\usepackage{multirow}
\usepackage{rotating}
\usepackage{tabularx}
\usepackage{gensymb}
\usepackage{graphicx}
\usepackage{float}
\usepackage{floatflt}
\usepackage{makecell}
\usepackage{colortbl}
\usepackage{calc}
\usepackage{subcaption}
\usepackage{booktabs}
\usepackage{arydshln}
\usepackage{amsmath}
\usepackage{bm}
\usepackage{enumitem}

\definecolor{cvprblue}{rgb}{0.21,0.49,0.74}
\usepackage[pagebackref,breaklinks,colorlinks,allcolors=cvprblue]{hyperref}

\usepackage[capitalize]{cleveref}
\crefname{section}{Sec.}{Secs.}
\Crefname{section}{Section}{Sections}
\Crefname{table}{Table}{Tables}
\crefname{table}{Tab.}{Tabs.}

\definecolor{colorFst}{HTML}{F59194}      
\definecolor{colorSnd}{HTML}{FAC791} 
\definecolor{colorThd}{HTML}{FFFF99}   

\newcommand{\defX}{\bm{x}}
\newcommand{\refX}{\mathbf{X}}
\newcommand{\dofs}{\mathbf{z}}

\newcommand{\R}{\mathbb{R}}
\newcommand{\W}{\bm{W}}
\newcommand{\dofsMat}{\mathbf{Z}}

\newcommand{\mass}{\mathbf{M}}
\newcommand{\NumHandle}{m}

\newcommand{\DeformationMap}{\phi}

\newcommand{\ElasticPotential}{E_\textrm{pot}}

\renewcommand{\eqref}[1]{Equation~\ref{eq:#1}}

\definecolor{myPurple}{rgb}{0.4, .0, .8}
\definecolor{myGreen}{rgb}{0, 0.6, .3}
\definecolor{myRed}{rgb}{0.8, .2, .2}
\definecolor{myOrange}{rgb}{0.8, 0.45, 0.0}
\definecolor{myBlue}{rgb}{.0, .0, 1.0}
\definecolor{myBlue2}{rgb}{.0, 1.0, 1.0}
\definecolor{myBlack}{rgb}{.0, .0, 0.0}
\definecolor{darkmidnightblue}{rgb}{0.0, 0.2, 0.4}
\definecolor{MyGreen}{rgb}{0.02,0.5,0.02}

\begin{document}

\setlength{\abovedisplayskip}{2.0pt} 
\setlength{\belowdisplayskip}{2.0pt}

\addtolength{\abovecaptionskip}{-1.0 em} 
\addtolength{\belowcaptionskip}{-1.0 em} 
\addtolength{\textfloatsep}{-0.75 em} 
\addtolength{\intextsep}{-0.5 em} 
\addtolength{\floatsep}{-0.75 em} 

\def\paperID{*****} 
\def\confName{CVPR}
\def\confYear{2026}

\title{
PhysSkin: Real-Time and Generalizable Physics-Based Animation \\ via Self-Supervised Neural Skinning
}

\author{
Yuanhang Lei$^{1}$\quad
Tao Cheng$^{1}$ \quad
Xingxuan Li$^{1}$  \quad
Boming Zhao$^{1}$ \quad \\
Siyuan Huang$^{2}$ \quad
Ruizhen Hu$^{3}$ \quad
Peter Yichen Chen$^{4}$ \quad
Hujun Bao$^{1}$ \quad
Zhaopeng Cui$^{1}$\footnotemark[2] \\
$^{1}$State Key Laboratory of CAD\&CG, Zhejiang University \quad
$^{2}$BIGAI \quad \\  $^{3}$Shenzhen University \quad $^{4}$University of British Columbia \quad 
}

\twocolumn[{
\renewcommand\twocolumn[1][]{##1}
\maketitle
\vspace{-3.5em}
\begin{center}
    \includegraphics[width=\textwidth]{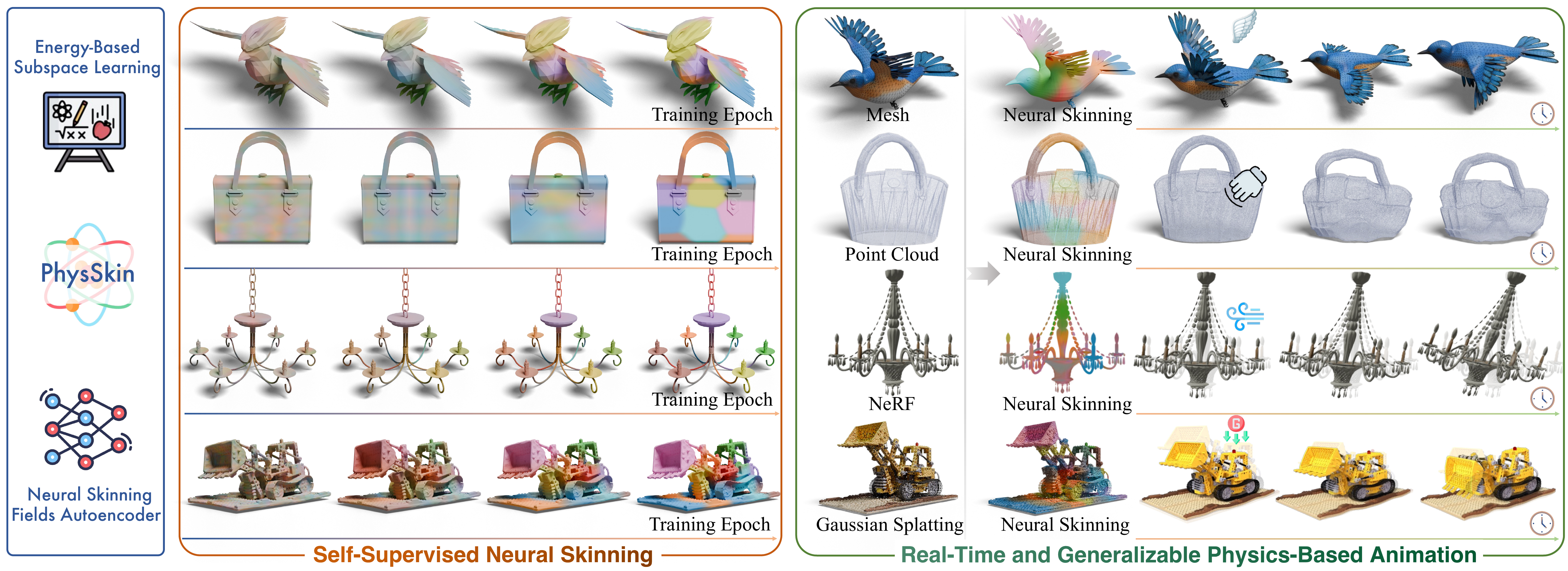}
    \captionof{figure}{PhysSkin is a generalizable physics-informed neural skinning framework for object animation. The framework is learned directly from static 3D geometries via physics-informed self-supervision without any annotated data. Once trained, PhysSkin can be applied in a feed-forward manner to perform neural skinning for diverse 3D shapes and discretizations, enabling real-time physics-based animation.}
    \label{fig:teaser}
\end{center}
\vspace{1.0em}
}]

\renewcommand{\thefootnote}{\fnsymbol{footnote}}
\footnotetext[2]{Corresponding author.}

\maketitle

\begin{abstract}
Achieving real-time physics-based animation that generalizes across diverse 3D shapes and discretizations remains a fundamental challenge. We introduce PhysSkin, a physics-informed framework that addresses this challenge. In the spirit of Linear Blend Skinning, we learn continuous skinning fields as basis functions lifting motion subspace coordinates to full-space deformation, with subspace defined by handle transformations. 
To generate mesh-free, discretization-agnostic, and physically consistent skinning fields that generalize well across diverse 3D shapes, PhysSkin employs a new neural skinning fields autoencoder which consists of a transformer-based encoder and a cross-attention decoder.
Furthermore, we also develop a novel physics-informed self-supervised learning strategy that incorporates on-the-fly skinning-field normalization and conflict-aware gradient correction, enabling effective balancing of energy minimization, spatial smoothness, and orthogonality constraints.
PhysSkin shows outstanding performance on generalizable neural skinning and enables real-time physics-based animation. Project Page: \url{https://zju3dv.github.io/PhysSkin/}.
\end{abstract}

\vspace{-0.5em}
\section{Introduction}
\label{sec:intro}

 \vspace{-0.5em}

\noindent Real-time physics-based animation is a long-standing goal in computer vision and graphics, underpinning applications in VR/AR authoring, character animation, and interactive digital content creation. To support such applications, an animation system must provide a compact yet expressive motion representation that captures physically plausible deformations and complex motion dynamics while remaining computationally efficient for real-time inference. 

Subspace physics-based animation~\cite{an2008optimizing, chen2023crom} addresses this by learning a low-dimensional motion subspace for efficient computation, and then lifting the result back to the full space using a subspace mapping. However, classical methods~\cite{an2008optimizing, sifakis2012fem} optimize a single linear subspace mapping matrix tied to a particular mesh topology and resolution, which prevents generalization across different spatial discretizations. Recently, neural methods~\cite{chen2023crom, sharp2023data} have emerged as promising alternatives by using neural networks to learn subspace mapping, which naturally handle arbitrary spatial discretizations, but these approaches must still train a separate network per object and fail to generalize to different 3D shapes. 
A complementary line of work defines subspace mapping through rigging and skinning. Recent neural rigging and skinning methods~\cite{xu2020rignet, deng2025anymate} can infer animation-ready skeletons and skinning weights directly from geometry, but they rely heavily on expert-annotated datasets and lack physical constraints, limiting their ability to model physically consistent deformations. These limitations motivate a central question: \emph{How can we learn a physics-consistent deformation subspace mapping that generalizes across diverse 3D shapes and discretizations directly from static geometries, without relying on expert-annotated data?}

In this work, we introduce PhysSkin, a generalizable physics-informed framework for real-time physics-based animation across diverse 3D shapes and discretizations. In the spirit of Linear Blend Skinning (LBS)~\cite{magnenat1989joint}, we learn continuous skinning fields as basis functions for deformation subspace mapping directly from static geometries via a new Neural Skinning Fields Autoencoder, with subspace coordinates defined by handle transformations. Unlike traditional approaches, PhysSkin requires no simulation trajectories, skinning annotations, or category-specific priors. The resulting representation is mesh-free, discretization-agnostic, and spatially continuous, enabling a single model to generalize across object categories, topologies, and discretization resolutions. Specifically, PhysSkin employs a transformer-based point cloud encoder to extract latent shape features, while a cross-attention decoder aggregates both surface and volumetric information via cubature points sampling, capturing the intrinsic structural cues necessary for physically plausible deformation and generating continuous neural skinning fields that generalize robustly to unseen shapes.

Learning such deformation subspace mapping purely from geometry is highly non-trivial. A physically meaningful subspace must exhibit:
(1) low potential energy, reflecting deformation modes compatible with physically plausible behavior;
(2) spatial smoothness, ensuring coherent, artifact-free deformations;
(3) orthogonality, providing numerically independent and stable eigenmodes.
However, these constraints often conflict in magnitude and direction, making naive joint optimization unstable and preventing prior methods from learning clean, physically consistent results.
To overcome this optimization challenge, we introduce a novel Physics-Informed Self-Supervised Learning (PISSL) strategy designed to enforce physical plausibility while ensuring numerical stability. First, we apply on-the-fly skinning-field normalization, which regulates the scale of the learned skinning weights and prevents numerical drift during training. Second, we incorporate a conflict-aware gradient correction mechanism~\cite{liu2025config} to resolve destructive interference among energy, smoothness, and orthogonality gradients, enabling balanced optimization and stable convergence. Together, these mechanisms allow PhysSkin to reliably discover clean, orthogonal, and physically consistent deformation subspace mapping, supporting real-time, physics-based animation across diverse 3D shapes.

Our contributions can be summarized as follows: (1) We present \textbf{PhysSkin}, a generalizable physics-informed framework for real-time physics-based animation across diverse 3D shapes and discretizations by learning neural skinning fields directly from static geometries; (2) We propose a new \textbf{Neural Skinning Fields Autoencoder} that combines a transformer-based shape encoder with a cross-attention decoder to produce mesh-free, discretization-agnostic, and spatially continuous skinning fields that are orthogonal and physically consistent; (3) We develop a novel \textbf{Physics-Informed Self-Supervised Learning (PISSL) strategy} that incorporates on-the-fly skinning-field normalization and conflict-aware gradient correction, enabling effective balancing of energy minimization, spatial smoothness, and orthogonality constraints; (4)  \textbf{Experiments} on various datasets demonstrate that PhysSkin achieves outstanding performance on generalizable neural skinning, while delivering real-time performance on physics-based animation.

\vspace{-0.8em}
\section{Related Work}
\vspace{-0.5em}
\noindent\textbf{Neural Subspace Physics-Based Animation.}
Subspace physics-based animation~\cite{an2008optimizing, von2013efficient, chen2023crom} is an effective method for accelerating animation computations. Its core idea is to learn a low-dimensional motion subspace for efficient computation, and then lifting the result back to the full space using a subspace mapping. Recently, fully neural approaches have been proposed to directly learn subspace mapping by neural networks. Data-driven methods~\cite{fulton2019latent, shen2021high, chen2023crom, zong2023neural, chang2023licrom} represented by  CROM~\cite{chen2023crom} learn the mapping for a single object by fitting neural networks to motion sequences from physical simulators, but they have limited generalization ability to unseen motion patterns. Data-free methods~\cite{sharp2023data, lyu2024accelerate, modi2024simplicits, wang2024neural, chang2025shape} learn a low-energy subspace mapping by physics-informed learning without requiring motion sequences, but pure neural network approaches often struggle to converge to high-quality subspace representations. Recently, Simplicits~\cite{modi2024simplicits} learns neural skinning weight functions as a physics-informed subspace mapping by minimizing a quadratic energy subject to orthogonality constraints, which stabilizes training compared to purely neural models but limits generalization due to its per-object network design. Our method overcomes these limitations with a neural skinning fields autoencoder that outputs skinning weights generalizing across diverse 3D shapes, enabling real-time subspace physics-based animation.

\noindent\textbf{Neural Rigging and Skinning Representations.}
Another classical reduced deformation representation is skeletal rigging-based animation~\cite{baran2007automatic, feng2015avatar, SMPL:2015}. Recently, based on 3D shape VAEs~\cite{zhang20233dshape2vecset, zhao2023michelangelo}, many works~\cite{deng2025anymate, song2025magicarticulate, guo2025make, liu2025riganything, xu2020rignet, zhang2025one, song2025puppeteer} have performed supervised learning from expert-annotated 3D object rigging and skinning datasets to infer objects' joints, bones, and skinning weights through a feed-forward network. However, some methods are limited to specific categories of 3D shapes such as characters~\cite{xu2020rignet, guo2025make} or animals~\cite{sun2025ponymation, wu2022casa}, as they incorporate category-specific priors~\cite{SMPL:2015, zuffi20173d} such as predefined skeletal structures or semantic constraints, making it difficult to generalize to other object categories. Works represented by Anymate~\cite{deng2025anymate} learn from annotated 3D object rigging and skinning datasets and design category-agnostic network architectures, thereby enabling the ability to generalize across diverse object categories and demonstrating stronger generalization capabilities. However, these methods heavily rely on expert-annotated data~\cite{deng2025anymate, li20214dcomplete, li2025puppet, wu2022casa, luo2023rabit}, which are costly and difficult to scale. In contrast, our work employs self-supervised learning to train neural skinning fields directly from static 3D geometries without any annotated data.

\begin{figure*}[t]
    \centering
    \vspace{-1.3em}
    \includegraphics[width=0.945\linewidth]{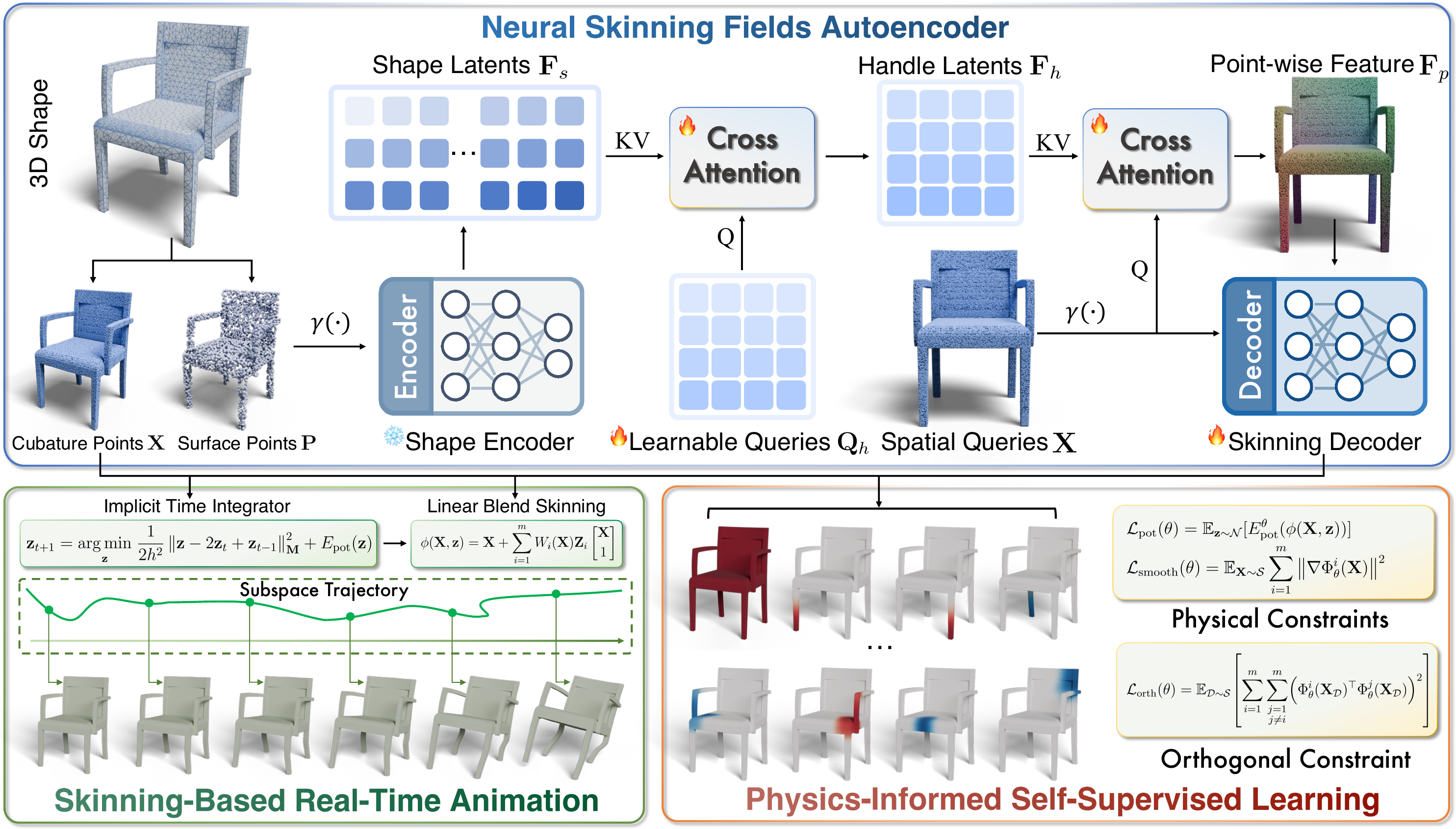}
    \caption{Given a static 3D shape, we first sample volumetric cubature points for animation and surface points for shape encoding. A shape encoder processes the surface points to produce shape latents, from which a set of learnable queries extract handle latents via cross-attention. Subsequently, spatial queries attend to the handle latents through another cross-attention to derive point-wise features, which are decoded into neural skinning fields. A geometry-only, physics-informed learning strategy optimizes the network in a self-supervised manner under physical and orthogonal constraints, enabling the learned skinning fields to support real-time, physics-based animation.}
    \vspace{-0.5em}
    \label{fig:pipeline}
\end{figure*}

\vspace{-0.8em}
\section{Method}
\vspace{-0.5em}
\noindent PhysSkin aims to construct a generalizable physics-informed framework for real-time physics-based animation across diverse 3D shapes and discretizations. As illustrated in Fig.~\ref{fig:pipeline}, our framework employs skinning-based representation to model full-space 3D object deformations within a reduced-order subspace, thereby enabling real-time physic-based animation(Sec.~\ref{subsec:skinning_eigenmodes}). To generalize the skinning-based subspace representation across different 3D shapes and various spatial discretizations, we propose a new neural skinning fields autoencoder that is mesh-free, discretization-agnostic, and spatially continuous(Sec.~\ref{subsec:skin_autoencoder}). Finally, to train our neural skinning fields autoencoder without any annotated data, we introduce a novel Physics-Informed Self-Supervised Learning (PISSL) strategy designed to enforce physical plausibility while ensuring numerical stability(Sec.~\ref{subsec:physics_informed_optimization}).

\vspace{-0.3em}
\subsection{Theory: Skinning-Based Real-Time Animation}
\vspace{-0.3em}
\label{subsec:skinning_eigenmodes}
\noindent\textbf{Full-Space Physics-Based Animation.}
Classical full-space physics-based animation methods~\cite{sifakis2012fem, liu2013fast, jiang2016material, xie2024physgaussian, lei2026diffwind} typically use explicit or implicit time integrators to update the object's motion state. Explicit time integrators are often constrained by stability conditions~\cite{li2025physics}, requiring multiple computational substeps per frame, which limits real-time animation capabilities. Implicit time integrators update the object's motion state through optimization, allowing for larger computational time step while maintaining stability~\cite{li2025physics}:
\vspace*{-0.2em}
\begin{equation}
    s_{t+1} =
    \underset{s}{\arg\min}\;
    \frac{1}{2h^2}\|s - 2s_t + s_{t-1}\|_{\mass}^2
    + \ElasticPotential(s),
    \label{eq:full_space_time_stepping}
\vspace{-0.2em}
\end{equation}
where $s \in \mathbb{R}^{3n}$ represents the full-space coordinates of the object with $n$ points, $h$ is the time step size, $\|\cdot\|_{\mass}$ is the metric induced by the mass matrix, and $\ElasticPotential$ is the potential energy that induces the internal force, with $\ElasticPotential$ supporting general hyperelastic material models. However, since it requires solving a large nonlinear optimization problem in the high-dimensional full space, implicit time integrators still struggle to achieve real-time physics-based animation.

\noindent\textbf{Skinning-Based Subspace.}
To model the full-space 3D object deformations within a reduced-order subspace, thereby enabling real-time physics-based animation, 
in the spirit of Linear Blend Skinning (LBS)~\cite{magnenat1989joint},  we represent the full-space 3D displacement field by modeling the deformation map $\defX = \DeformationMap(\refX, \dofs(t))$ as a weighted sum of $m$ affine transformations applied to the rest-pose positions as FastCody~\cite{benchekroun2023fast}:
\vspace*{-0.5em}
\begin{equation} \label{eq:lbs_equation}
   \DeformationMap(\refX, \dofs) = \refX + \sum_{i=1}^\NumHandle \W_{\!i} (\refX) \dofsMat_i  \begin{bmatrix} \refX \\ 1 \end{bmatrix},
\vspace{-0.4em}
\end{equation}
where $\refX \in \R^3$ is a point in undeformed space and $\defX\in\R^3$ is its deformed position according to $\dofs(t)$, $\NumHandle$ is the number of \emph{skinning handles}, $\W_{\!i}(\refX) \in \R^3 \rightarrow \R$ is the skinning weight function associated with handle $i$, producing signed weights indicating the influence of handle $i$, $\dofsMat_i\in\R^{3\times4}$ is the $i^{th}$ skinning handle, $\dofs=\text{flat}(\dofsMat) \in \R^{12m}$ denotes the flattened vector of all stacked handle transformations. In this formulation, the vector $\dofs$ corresponds to the reduced-order subspace coordinates, allowing the object's full deformation in the high-dimensional space $s \in \mathbb{R}^{3n}$ to be represented compactly and efficiently, where $m \ll n$.

\noindent\textbf{Subspace Dynamics.}
With the skinning-based subspace representation, we can create realistic motions with physically plausible deformations directly in the reduced-order subspace. 
We discretize the governing Newtonian physical equations using standard implicit time integration in Eq.~(\ref{eq:full_space_time_stepping}) and update the subspace coordinates $\dofs$ over time as:
{\small
\vspace*{-0.6em}
\begin{equation}
    \dofs_{t+1} =
    \underset{\dofs}{\arg\min}\;
    \frac{1}{2h^2}\|\dofs - 2\dofs_t + \dofs_{t-1}\|_{\mass}^2
    + \ElasticPotential(\DeformationMap(\refX, \dofs)),
    \label{eq:time_stepping}
\vspace{-0.6em}
\end{equation}}
where $h$, $\|\cdot\|_{\mass}$, and $\ElasticPotential$  are the same as defined in Eq.~(\ref{eq:full_space_time_stepping}). This formulation can be solved using standard Newton-based methods~\cite{nocedal2006numerical}.
Since the dimensionality of the reduced subspace is significantly smaller than that of the full space, 
the optimization in Eq.~(\ref{eq:time_stepping}) converges rapidly~\cite{benchekroun2023fast}, thereby enabling real-time physics-based animation.

\vspace{-0.3em}
\subsection{Neural Skinning Fields Autoencoder}
\label{subsec:skin_autoencoder}
\vspace{-0.3em}
\noindent Our goal is to learn a mesh-free, discretization-agnostic, and spatially continuous neural skinning fields autoencoder that generalizes across different 3D shapes and discretizations for skinning-based real-time animation in Sec.~\ref{subsec:skinning_eigenmodes}. Unlike existing neural subspace methods~\cite{modi2024simplicits, chen2023crom} that train a separate neural network for each individual object, we aim to train a generalizable feed-forward neural network that can be applied to a wide range of objects, which significantly enhances the network's scalability and practicality.

\noindent\textbf{Neural 3D Shape Encoding.}
We use  transformer-based point cloud encoder Michelangelo~\cite{zhao2023michelangelo} to extract a highly compressed latent set $\mathbf{F}_s \in \mathbb{R}^{M \times d}$ for given 3D shape via cross-attention block and several self-attention blocks:
\begin{equation}
\mathbf{F}_s = \mathrm{SelfAttn}^{(1:L)}(\mathrm{CrossAttn}(\mathbf{Q}_s, \gamma(\mathbf{P}))),
\label{eq:3dshape_encoder}
\end{equation}
where $\mathbf{Q}_s \in \mathbb{R}^{M \times d}$ is a set of learnable shape latent tokens, $\mathbf{P} \in \mathbb{R}^{N \times 6}$ is the sampled surface point cloud with $N$ points, each point consisting of 3D coordinates and normals, $\gamma(\cdot)$ is point-wise positional encoding , $\mathrm{SelfAttn}^{(1:L)}(\cdot)$ indicates a stack of $L$ self-attention layers that iteratively improve the shape latent representation, $M$ is the number of latents, and $d$ is the dimension of each latent.
In practice, the shape encoder is pretrained on ShapeNet dataset~\cite{chang2015shapenet} via a signed distance field (SDF) reconstruction task~\cite{park2019deepsdf} to learn generalizable geometric priors. We then freeze the encoder parameters during training to enhance efficiency.

\noindent\textbf{Neural Skinning Fields Decoding.}
To represent a generalizable, discretization-agnostic, and spatially continuous neural skinning fields decoder, we decode the shape latent set $\mathbf{F}_s$ obtained from Eq.~(\ref{eq:3dshape_encoder}) into spatially varying skinning fields across different 3D shapes.
We first assign $m$ learnable skinning handle tokens $\mathbf{Q}_h \in \mathbb{R}^{m \times d}$, and then use a cross-attention block to extract skinning handle latent set $\mathbf{F}_h \in \mathbb{R}^{m \times d}$ from the shape latent set $\mathbf{F}_s$:
\vspace*{0.3em}
\begin{equation}
\mathbf{F}_h = \mathrm{CrossAttn}(\mathbf{Q}_h, \mathrm{SelfAttn}(\mathbf{F}_s)).
\label{eq:skinning_handle_feature}
\vspace{0.3em}
\end{equation}
Next, for any spatial query point $\refX$, we use another cross-attention block to extract point-wise skinning feature $\mathbf{F}_p \in \mathbb{R}^{d}$ from the skinning handle latent set $\mathbf{F}_h$:
\vspace*{0.3em}
\begin{equation}
\mathbf{F}_p = \mathrm{CrossAttn}(\gamma(\refX), \mathbf{F}_h).
\label{eq:pointwise_skinning_feature}
\vspace{0.3em}
\end{equation}
Finally, we concatenate the point-wise skinning feature $\mathbf{F}_p$ with the positional encoding of the query point, and use a ResNet-style MLP to compute the skinning fields:
\vspace*{0.2em}
\begin{equation}
\W(\refX) = \mathrm{MLP}(\mathbf{F}_p \oplus \gamma(\refX)),
\label{eq:skinning_weights_decoder}
\vspace{0.2em}
\end{equation}
where $\W(\refX) \in \mathbb{R}^{m}$ is the skinning vector at point $\refX$, and $\oplus$ denotes the concatenation operator. To promote the orthogonality among different skinning modes, we apply Orthogonalization by Newton's Iteration (ONI) module~\cite{huang2020controllable} at the final layer of the MLP, we adopt ELU as the activation function in the MLP, which allows learning flexible and expressive skinning-based deformation subspaces without enforcing non-negativity on the learned skinning modes. We denote our full neural skinning fields autoencoder as $\Phi_\theta(\refX)$.

\noindent\textbf{Cubature Points Sampling.}
In classical Finite Element Method (FEM)~\cite{sifakis2012fem}, the object's volume is typically discretized into a tetrahedral mesh~\cite{hu2020fast} to facilitate numerical solutions of physical equations.
Unlike triangle surface meshes commonly used for rendering, which only contain surface points, tetrahedral meshes include interior volumetric elements that better capture the object's spatial structure. To learn a discretization-agnostic and spatially continuous neural skinning fields autoencoder, we sample cubature points both on object's surface and inside its volume, each cubature point serves as a spatial query point $\refX$, which is further used to compute the $\ElasticPotential$ in Eq.~(\ref{eq:full_space_time_stepping}) and Eq.~(\ref{eq:time_stepping}), and thus capture volumetric deformation behavior that surface-only samplers cannot, this contrasts with popular 3D Shape VAEs~\cite{zhang20233dshape2vecset} that only sample surface points. To sample interior cubature points, we first convert the object's surface mesh into a watertight mesh~\cite{huang2018robust, huang2020manifoldplus}, we then uniformly sample points in the voxelized spatial grid using ray tracing~\cite{3dgrt2024}, and classify points as being outside the object if the ray intersects the surface mesh and the dot product with the normal is negative. For surface cubature points, we employ Sharp Edge Sampling (SES)~\cite{chen2025dora} to better capture geometric details.

\begin{figure}[t]
    \centering
    \includegraphics[width=1.00\linewidth]{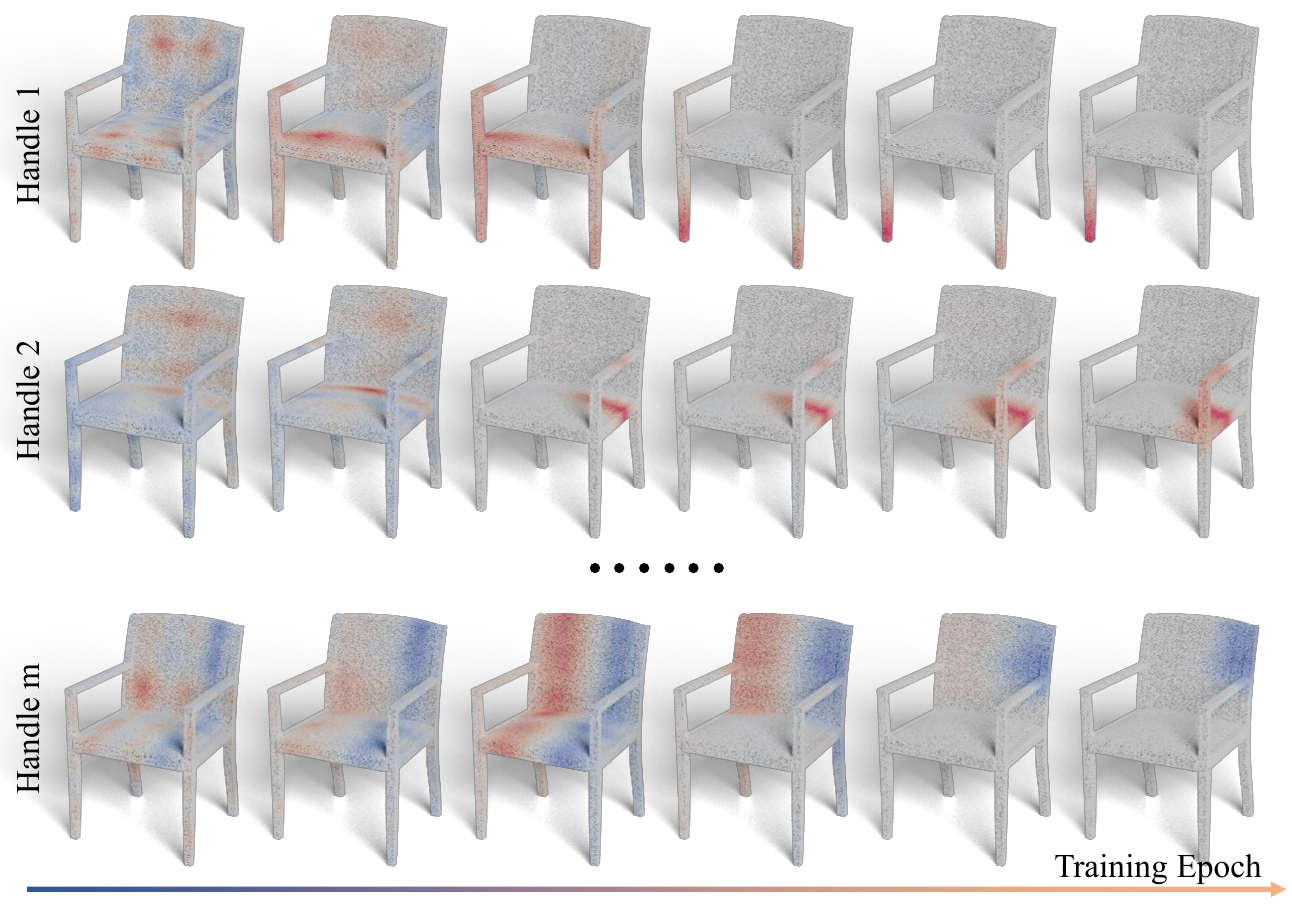}
    \caption{Evolution of neural skinning fields during optimization. 
Starting from disordered initial representations, 
our physics-informed self-supervised learning progressively organizes them into physically consistent, geometrically orthogonal, and spatially smooth skinning results. Each skinning weight $i$ is scaled by $\max(abs(\W_i))$ to fall within $[-1, 1]$ and centered around 0.}
    \vspace{-0.2em}
    \label{fig:ours_training_progress}
\end{figure}

\vspace{-0.2em}
\subsection{Physics-Informed Self-Supervised Learning}
\vspace{-0.2em}
\noindent Current popular feed-forward neural skinning methods~\cite{deng2025anymate} typically require expert-annotated skinning data for supervised learning, which is often obtained through complex manual annotation~\cite{baran2007automatic}, making the data acquisition costly. Additionally, these models do not inherently understand the underlying physical principles.
Moreover, most methods are limited to rigging for animals or characters as they incorporate category-specific priors and do not generalize well to general object categories. To address these limitations, we design a Physics-Informed Self-Supervised Learning (PISSL) strategy for training our neural skinning fields autoencoder to improve generalization and practicality.

\label{subsec:physics_informed_optimization}
\noindent\textbf{Physical Constraints.}
To ensure that the output of $\Phi_\theta(\refX)$ is physically plausible, we should construct a low-energy subspace~\cite{sharp2023data} by minimizing the expected potential energy of randomly sampled subspace coordinates $\dofs \in \R^{12m}$:
\vspace*{0.4em}
\begin{equation}
\mathcal{L}_{\text{pot}}(\theta) = \mathbb{E}_{\dofs \sim \mathcal{N}}[\ElasticPotential^{\theta}(\DeformationMap(\refX, \dofs))],
\label{eq:energy_loss}
\vspace{0.4em}
\end{equation}
where $\dofs$ is defined in Sec.~\ref{subsec:skinning_eigenmodes}, $\mathcal{N}$ denotes a Gaussian distribution over $\mathbb{R}^{12m}$ with zero mean and variance $\mu^2 I_{d\times d}$ with $\mu=0.2$, $\mathbb{E}$ denotes the average potential energy over the sampled subspace coordinates $\dofs$ within a training batch, and $\ElasticPotential^{\theta}(\dofs)$ is the potential energy computed under the current network parameters $\theta$. During training, we linearly interpolate the potential energy computation from linear elasticity~\cite{teran2005robust} model to Neo-Hookean~\cite{kim2020dynamic} model to enhance training stability~\cite{modi2024simplicits}. Besides the low-energy constraint imposed by $\mathcal{L}_{\text{pot}}$, we also expect the output of $\Phi_\theta(\refX)$ to be spatially smooth, so we introduce a spatial smoothness loss:
\vspace*{-0.2em}
\begin{equation}
\mathcal{L}_{\text{smooth}}(\theta) = \mathbb{E}_{\refX \sim \mathcal{S}} \sum_{i=1}^{m} \|\nabla \Phi_{\theta}^{i}(\refX)\|^2,
\vspace{-0.2em}
\label{eq:smoothness_loss}
\end{equation}
where $\mathcal{S}$ denotes the set of sampled cubature points introduced in Sec.~\ref{subsec:skin_autoencoder}, and $\nabla \Phi_{\theta}^{i}$ is the spatial gradient of the $i^{th}$ skinning weight function at point $\refX$ under the current network parameters $\theta$. Through the physics-informed self-supervised learning, we can train our neural skinning fields autoencoder without any annotated skinning data or precomputed simulation trajectories, significantly enhancing the model's practicality. Additionally, the model can perceive the physical principles of the underlying mechanical system.

\noindent\textbf{Orthogonal Constraint.}
Besides physical constraints, we also expect different skinning weight functions to be orthogonal to enhance the expressiveness of the subspace representation. Therefore, we introduce an orthogonal loss:
{\small
\vspace*{-0.3em}
\begin{equation}
\mathcal{L}_{\mathrm{orth}}(\theta)
= \mathbb{E}_{\mathcal{D}\sim\mathcal{S}}
\left[
\sum_{i=1}^{m} \sum_{\substack{j=1\\ j\neq i}}^{m}
\Big( \Phi_{\theta}^{i}(\mathbf{X}_{\mathcal{D}})^\top \Phi_{\theta}^{j}(\mathbf{X}_{\mathcal{D}}) \Big)^2
\right],
\vspace{-0.3em}
\label{eq:orth_sigma_complete}
\end{equation}
}
where $\mathcal{D}$ denotes a training batch of cubature points sampled from the overall set $\mathcal{S}$, and $\mathbf{X}_{\mathcal{D}}$ represents the corresponding point matrix within the batch. Unlike Simplicits~\cite{modi2024simplicits} which simultaneously constrains the magnitude of skinning weights to be 1 in the orthogonal constraint process, making it difficult for the orthogonality constraint to converge and leading to unstable training gradients, we on-the-fly $\ell_2$-normalize each column of the skinning modes matrix, which regulates the scale of the learned modes and prevents numerical drift during training, this facilitates the convergence of the orthogonality constraint, leading to a more independent and well-conditioned neural skinning fields. Our overall loss function for physics-informed self-supervised learning is:
\vspace*{0.2em}
\begin{equation}
\mathcal{L}(\theta) = \mathcal{L}_{\text{smooth}}(\theta) + \lambda_{\text{pot}} \mathcal{L}_{\text{pot}}(\theta) + \lambda_{\text{orth}} \mathcal{L}_{\text{orth}}(\theta),
\label{eq:overall_loss}
\vspace{0.2em}
\end{equation}
where $\lambda_{\text{pot}}$ and $\lambda_{\text{orth}}$ are the weights for the potential energy loss and orthogonal loss. 
However, these constraints often conflict in magnitude and direction, making naive joint optimization unstable and preventing prior methods~\cite{sharp2023data, modi2024simplicits} from learning clean, physically consistent results. To overcome this challenge, we complement our on-the-fly skinning-field normalization with ConFIG~\cite{liu2025config} to correct conflicting gradients, resulting in stable and conflict-free optimization.
In Fig.~\ref{fig:ours_training_progress}, we visualize the evolution of different skinning modes during optimization, demonstrating that our physics-informed self-supervised framework progressively organizes the initially unstructured skinning representations into physically consistent, geometrically orthogonal, and spatially smooth skinning results, which are then used for real-time physics-based animation as detailed in Sec.~\ref{subsec:skinning_eigenmodes}.

\section{Experiments}
\label{sec:experiments}
\vspace{-0.2em}
\subsection{Implementation Details}
\label{subsec:Implementation}
\vspace{-0.2em}
\noindent For the 3D shape encoder, we adopt the pre-trained Michelangelo~\cite{zhao2023michelangelo} model which consists of one cross-attention block and eight self-attention blocks, the output latent set $\mathbf{F}_s \in \mathbb{R}^{M \times d}$ has $M = 256$ and $d=768$. For each 3D object, we re-center and normalize it into a bounding volume of $[-1, 1]^3$. For the 3D shape encoder, we sample 4096 points on the object surface as input. For cubature points sampling, we first sample 100k points on the surface, and then voxelize the space into a $128^3$ grid to obtain interior points to form the complete candidate cubature point set. For each training batch, we randomly sample 1000 cubature points from the candidate point set for potential energy computation, following the same strategy as previous work~\cite{chang2023licrom, modi2024simplicits}. In each batch, we randomly sample 1024 subspace coordinates $\dofs$ for training. Our framework is trained on 4 NVIDIA GeForce RTX 4090 GPUs. The learning rate is linearly increased to $5\times 10^{-4}$ within the first $1\%$ iterations (warm-up), and then gradually decreased using the cosine decay schedule until reaching the minimum value of $5\times10^{-5}$.

\vspace{-0.2em}
\subsection{Evaluation of Learned Skinning Fields}
\vspace{-0.2em}
\noindent\textbf{Baselines.}
We compare our method with the following baselines: Simplicits~\cite{modi2024simplicits}, RigNet~\cite{xu2020rignet}, M-I-A~\cite{guo2025make}, Anymate~\cite{deng2025anymate}, and Puppeteer~\cite{song2025puppeteer} on ShapeNet~\cite{chang2015shapenet} and RigNet~\cite{xu2020rignet} datasets. Simplicits train a separate neural network for each object to output skinning fields through self-supervised learning. Other baselines are supervised neural rigging and skinning methods that output object joints, bones, and skinning weights in a feed-forward manner, these methods are pretrained on expert-annotated datasets before being fine-tuned and evaluated with respect to our approach.

\noindent\textbf{Evaluation Metrics.}
Since our framework is entirely self-supervised and does not rely on annotated skinning data as `ground truth'
for metric evaluation like supervised approaches, we draw from matrix analysis~\cite{horn1994topics, axler2024linear} and spectral theory~\cite{stoica2005spectral, de2016spectral} to propose three quantitative metrics that measure \emph{structural independence},
\emph{numerical stability}, and \emph{spectral balance} of the skinning weights matrix $W \in \mathbb{R}^{N\times K}$, where $N$ is the number of vertices and $K$ is the number of handles or bones. 
To ensure a fair comparison, we perform $\ell_2$-normalization on each column of \(W\) before evaluation for all methods, resulting in the matrix \(\hat{W}\). 
 
\noindent \textbf{(1) Orthogonality Metric.}
We first measure the pairwise orthogonality among handle or bone influence vectors as:
\vspace*{-0.1em}
\begin{equation}
\Omega_{\text{orth}} = 
\frac{1}{K(K-1)} \|\,\hat{W}^\top \hat{W} - I\,\|_2^2,
\vspace{-0.1em}
\end{equation}
where $I$ is the $K\times K$ identity matrix. A smaller $\Omega_{\text{orth}}$ indicates that different handle or bone weights are more decorrelated and capture distinct deformation subspaces. Since each column of $\hat{W}$ is $\ell_2$-normalized, $\Omega_{\text{orth}}$ is bounded within $[0, 1]$, 
and a smaller value indicates that each handle or bone encodes an independent deformation mode.

\begin{table}[t]
\centering
\caption{
Quantitative comparison on RigNet~\cite{xu2020rignet} dataset.
}
\resizebox{0.95\linewidth}{!}{
\tabcolsep 8pt
\footnotesize
\begin{tabular}{lccc}
\toprule
\multicolumn{1}{c}{Method} & \multicolumn{1}{l}{$\Omega_{\text{orth}} \times 10^{-2}$ $\downarrow$} & \multicolumn{1}{l}{$\kappa_{\log}$ $\downarrow$} & \multicolumn{1}{l}{$H_{\text{spec}}$ $\uparrow$}  \\ 
\midrule
RigNet~\cite{xu2020rignet} & \cellcolor[HTML]{FAC791}0.5324 & \cellcolor[HTML]{FFFF99}2.7997  & \cellcolor[HTML]{FFFF99}0.9762 \\
M-I-A~\cite{guo2025make} & 1.4098  & 27.7357  & 0.7224  \\
Anymate~\cite{deng2025anymate} & 1.5737  & \cellcolor[HTML]{FAC791}2.6093 & 0.9682  \\
Puppeteer~\cite{song2025puppeteer} & \cellcolor[HTML]{FFFF99}0.5615  & 5.5605 & \cellcolor[HTML]{FAC791}0.9798 \\
Ours & \cellcolor[HTML]{F59194}0.0033 & \cellcolor[HTML]{F59194}1.0453 & \cellcolor[HTML]{F59194}0.9999   \\
\bottomrule
\end{tabular}
}
\vspace{-1.0em}
\label{tab:ours_vs_baselines_unseenobj}
\end{table}

\begin{figure}[!t]
    \centering
    \vspace{0.6em}
    \includegraphics[width=0.95\linewidth]{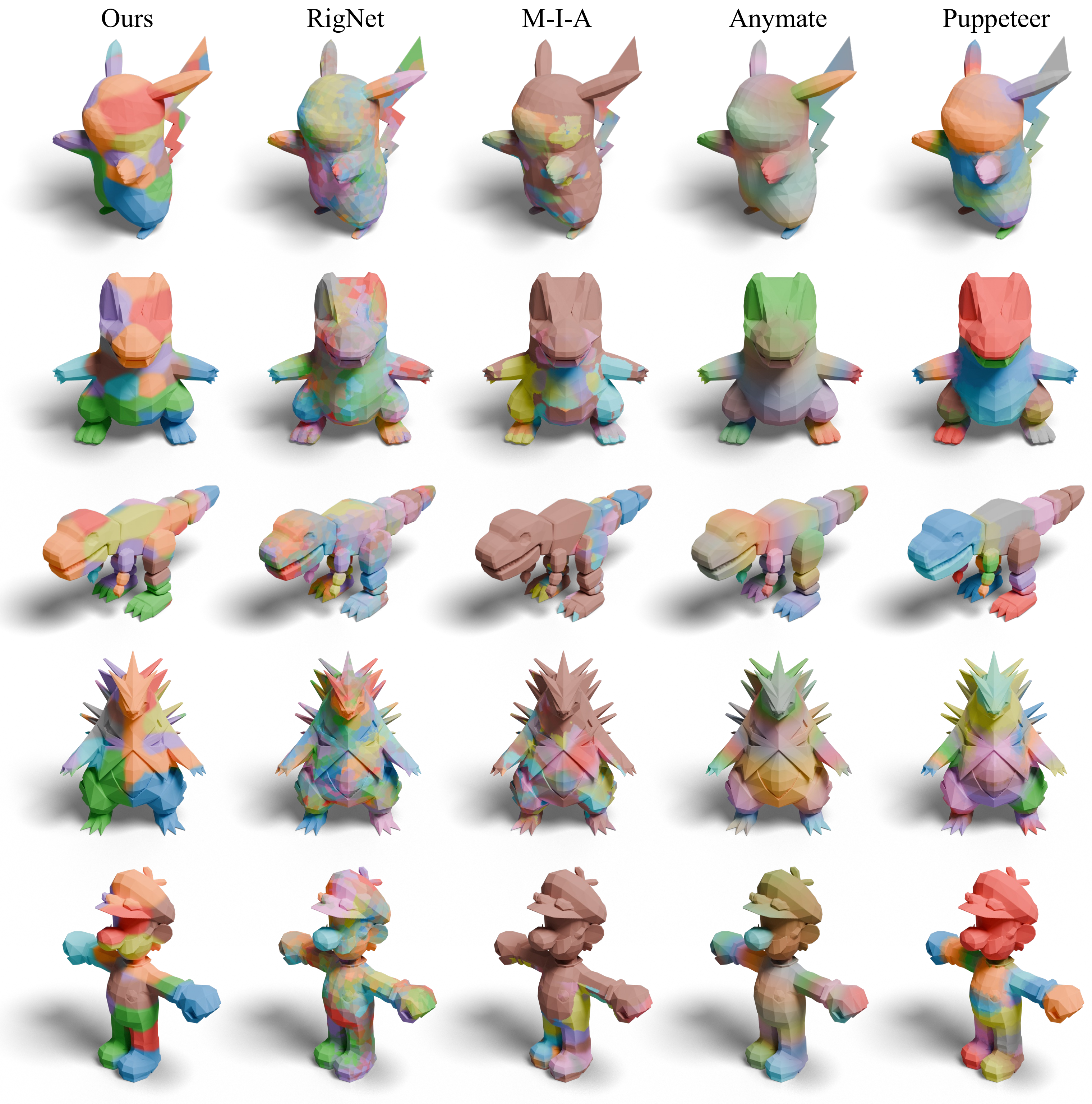}
    \caption{Qualitative comparisons on RigNet~\cite{xu2020rignet} dataset. We visualize combined skinning fields obtained by blending all skinning weights together, providing an overall view of deformation influences and smoothness compared with baselines~\cite{xu2020rignet, guo2025make, deng2025anymate, song2025puppeteer}.}
    \vspace{-0.5em}
    \label{fig:rignet-baselines}
\end{figure}

\noindent \textbf{(2) Log-Condition Number Metric.} 
To evaluate the numerical stability of the skinning basis, we compute the log-scaled  condition number~\cite{higham1995condition} of the Gram matrix:
\vspace*{-0.1em}
\begin{equation}
\kappa_{\log} = \log_2 \Big( 1 + \frac{\lambda_{\max}(\hat{W}^\top \hat{W})}{\lambda_{\min}(\hat{W}^\top \hat{W})} \Big),
\vspace{-0.1em}
\end{equation}
where $\lambda_{\max}$ and $\lambda_{\min}$ are the largest and smallest eigenvalues of $\hat{W}^\top \hat{W}$. 
$\kappa_{\log}$ lies within $[1, +\infty)$, a smaller value indicates a well-conditioned deformation basis, where small transformations result in stable geometric responses.

\noindent \textbf{(3) Spectrum Entropy Metric.} To assess the balance of spectral energy distribution across skinning weights, we define the spectrum entropy metric~\cite{de2016spectral, stoica2005spectral} as:
\vspace*{-0.1em}
\begin{equation}
H_{\text{spec}}  =
- \frac{1}{\log K}\sum_{i=1}^{K}
p_i \log p_i, 
\;
p_i = \frac{\lambda_i}{\sum_j \lambda_j},
\vspace{-0.1em}
\end{equation}
where $\{\lambda_i\}_{i=1}^{K}$ are the eigenvalues of $\hat{W}^\top \hat{W}$. $H_{\text{spec}}$ is bounded within $[0, 1]$, where
higher values correspond to more uniform spectral energy distribution, reflecting a balanced and expressive deformation modes.

\begin{table}[t!]
\centering
\caption{
Quantitative comparison on ShapeNet dataset~\cite{chang2015shapenet}.
}
\resizebox{0.88\linewidth}{!}{
\tabcolsep 8pt
\footnotesize
\begin{tabular}{lccc}
\toprule
\multicolumn{1}{c}{Method} & \multicolumn{1}{l}{$\Omega_{\text{orth}} \times 10^{-2}$ $\downarrow$ } & \multicolumn{1}{l}{$\kappa_{\log}$ $\downarrow$} & \multicolumn{1}{l}{$H_{\text{spec}}$ $\uparrow$}  \\ 
\midrule
RigNet~\cite{xu2020rignet} & \cellcolor[HTML]{FFFF99}0.5130 & 4.5417  & 0.9648  \\
M-I-A~\cite{guo2025make} & 1.5736 & 27.9573  & 0.7006  \\
Anymate~\cite{deng2025anymate} & 5.3520 & 4.9221   & 0.8858  \\
Puppeteer~\cite{song2025puppeteer} & 1.9528  & \cellcolor[HTML]{FAC791}1.4317  & \cellcolor[HTML]{FFFF99}0.9799 \\
Simplicits~\cite{modi2024simplicits} & \cellcolor[HTML]{FAC791}0.2621 & \cellcolor[HTML]{FFFF99}1.5205 & \cellcolor[HTML]{FAC791}0.9941 \\
Ours & \cellcolor[HTML]{F59194}0.0098 & \cellcolor[HTML]{F59194}1.0460 & \cellcolor[HTML]{F59194} 0.9997  \\
\bottomrule
\end{tabular}
}
\vspace{-1.0em}
\label{tab:ours_vs_baselines_seenobj}
\end{table}

\begin{figure}[t]
    \centering
    \vspace{0.35em}
    \includegraphics[width=0.88\linewidth]{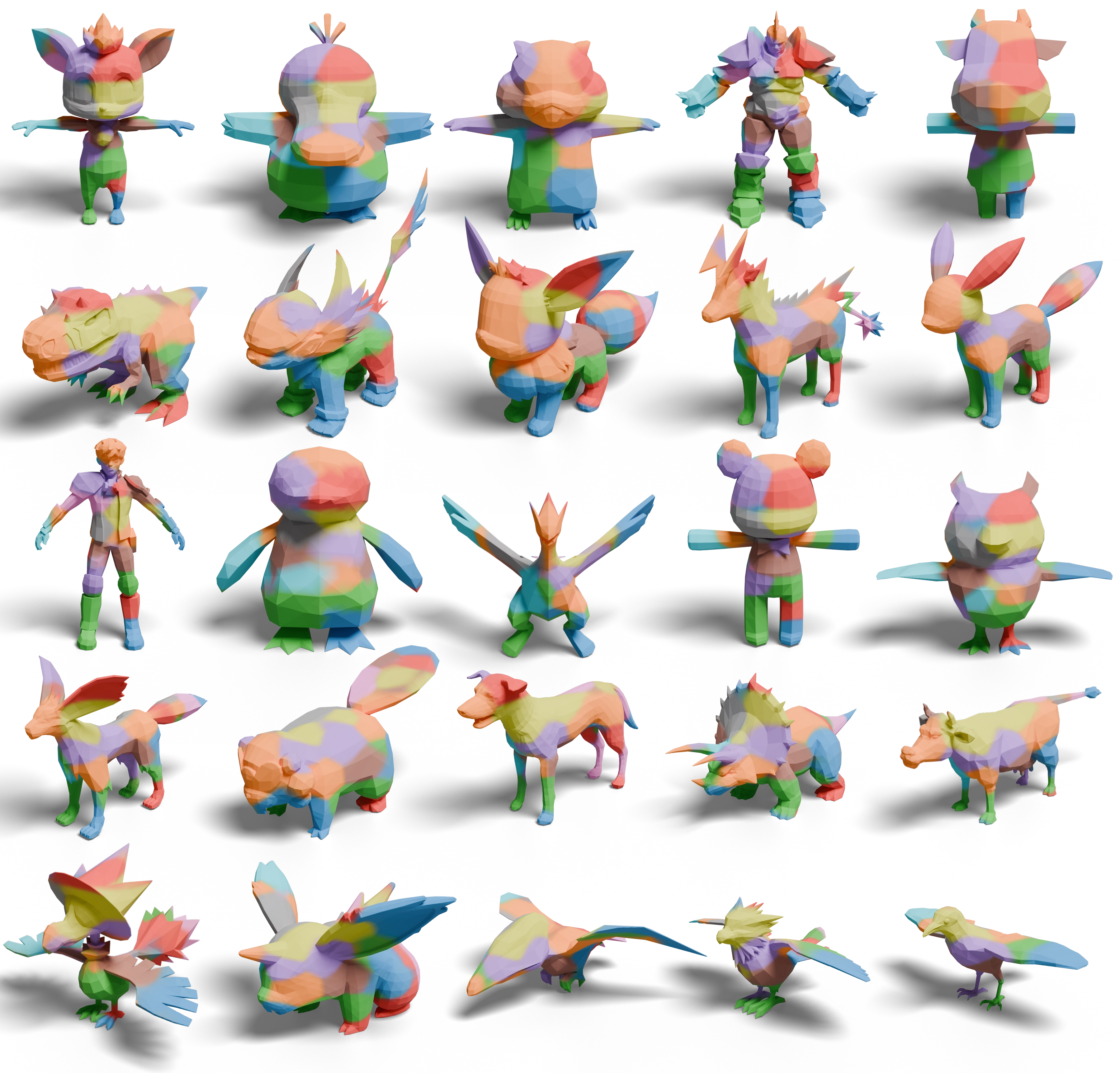}
    \vspace{0.1em}
    \caption{Qualitative results on unseen 3D shapes from the RigNet~\cite{xu2020rignet} test set. We visualize blended skinning fields to demonstrate our method's generalization to novel 3D shapes. \textbf{All results are produced by a single unified PhysSkin model.}}
    \label{fig:rignet-unseen}
\end{figure}

\begin{figure}[t]
    \centering
    \vspace{-1.0em}
    \includegraphics[width=0.95\linewidth]{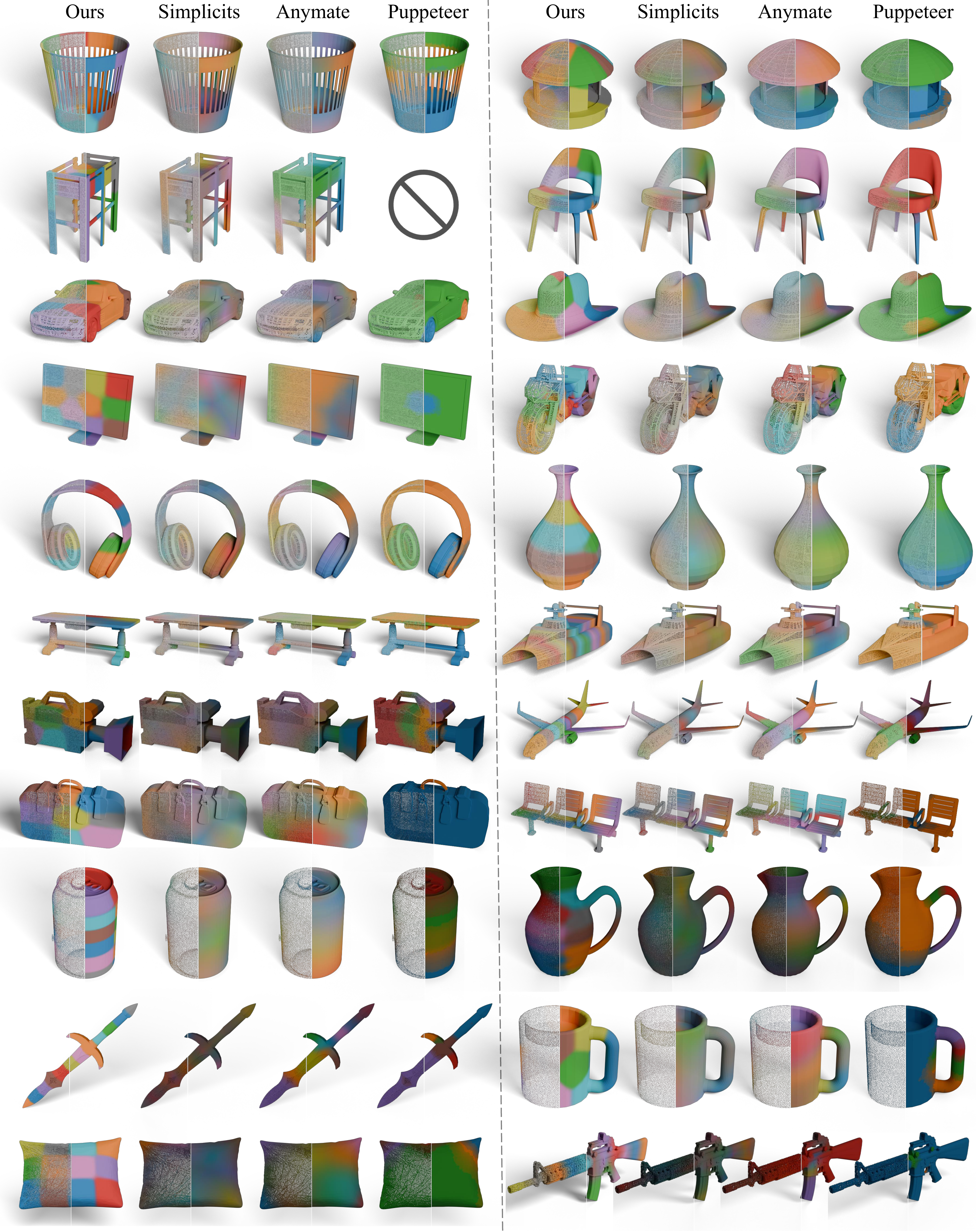}
    \caption{Qualitative comparisons on ShapeNet~\cite{chang2015shapenet} dataset. We compare our methods with Simplicits~\cite{modi2024simplicits}, Anymate~\cite{deng2025anymate}, and Puppeteer~\cite{song2025puppeteer}. \textbf{Zoom in} for more details.}
    \vspace{0.5em}
    \label{fig:ours_vs_baselines_shapenet}
\end{figure}

\begin{figure}[t]
    \centering
    \vspace{-0.1em}
    \includegraphics[width=0.95\linewidth]{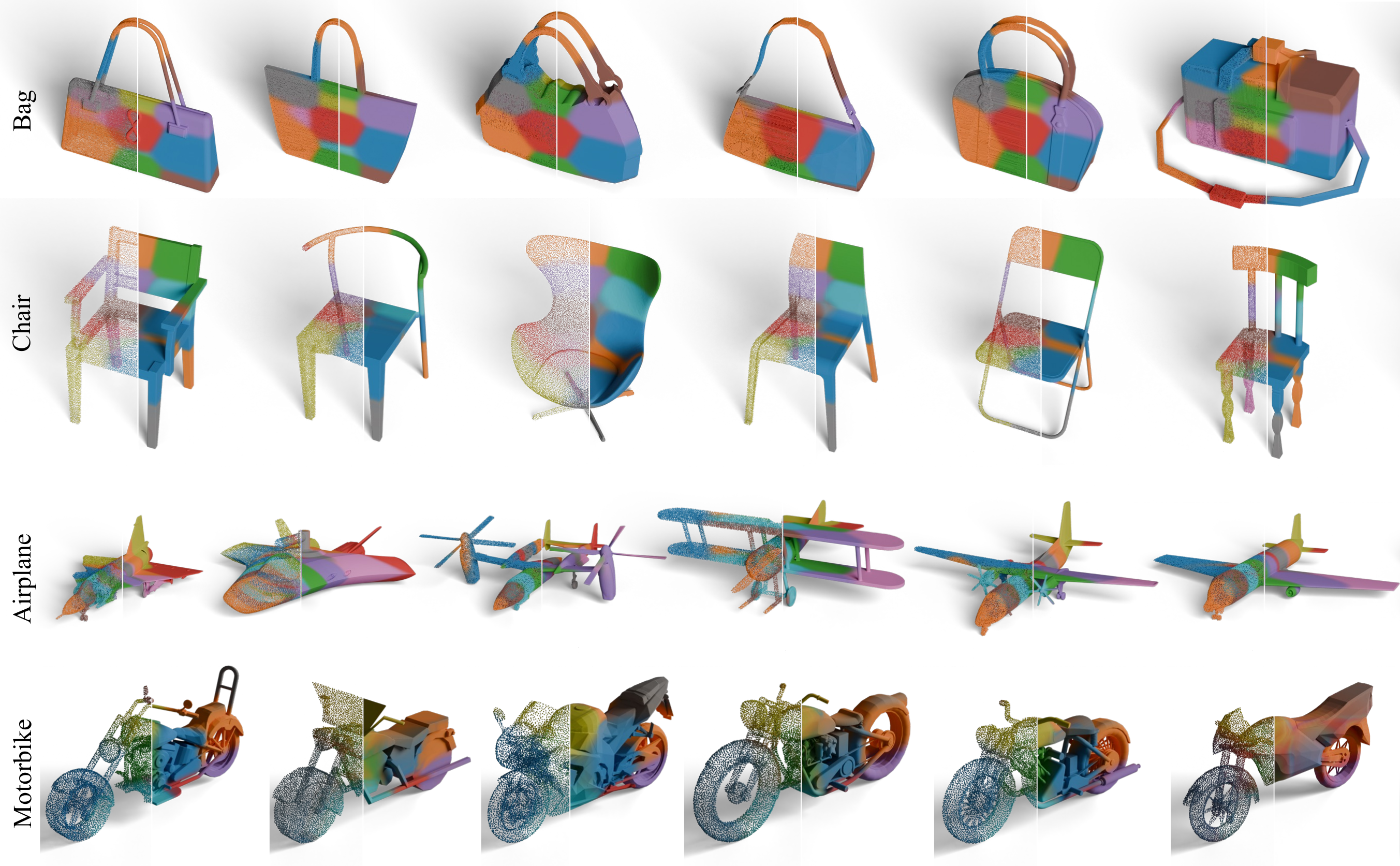}
    \vspace{-0.1em}
    \caption{Qualitative skinning results of our method on unseen objects from the ShapeNet~\cite{chang2015shapenet} dataset. \textbf{Zoom in} for more details.}
    \vspace{0.4em}
    \label{fig:ours_unseen}
\end{figure}

\noindent\textbf{Results.}We first evaluate our method on neural skinning for unseen objects from the test set of the RigNet~\cite{xu2020rignet} dataset to assess its generalization ability. RigNet dataset provides annotated skinning data, and we fine-tune the pre-trained models of feed-forward neural skinning baselines and our model is trained only on the 3D meshes. We present the quantitative comparison results on unseen objects in Table~\ref{tab:ours_vs_baselines_unseenobj}. Our method significantly outperforms the baselines on all three evaluation metrics,
demonstrating the generalization ability of our method.
We further evaluate our method on the ShapeNet dataset~\cite{chang2015shapenet}, which provides clearly defined object categories, our model is trained across all categories, and all baselines are evaluated on the corresponding test sets. Because Simplicits can only be trained on a single object at a time, we train an individual Simplicits model for each object. The quantitative results in Table~\ref{tab:ours_vs_baselines_seenobj} show that our method consistently outperforms all baselines.
In Fig.~\ref{fig:rignet-baselines} and Fig.~\ref{fig:ours_vs_baselines_shapenet}, we show the qualitative skinning results compared with the baselines, demonstrating that our method can obtain more physically consistent, geometrically orthogonal, and spatially smooth skinning results. In Fig.~\ref{fig:rignet-unseen} and Fig.~\ref{fig:ours_unseen}, we show the qualitative results of our method on unseen objects, demonstrating that our method can generalize well to unseen objects and obtain high-quality skinning results. For more skinning results, please refer to the supplementary.

\begin{figure}[t]
    \centering
    \vspace{-0.2em}
    \includegraphics[width=0.9\linewidth]{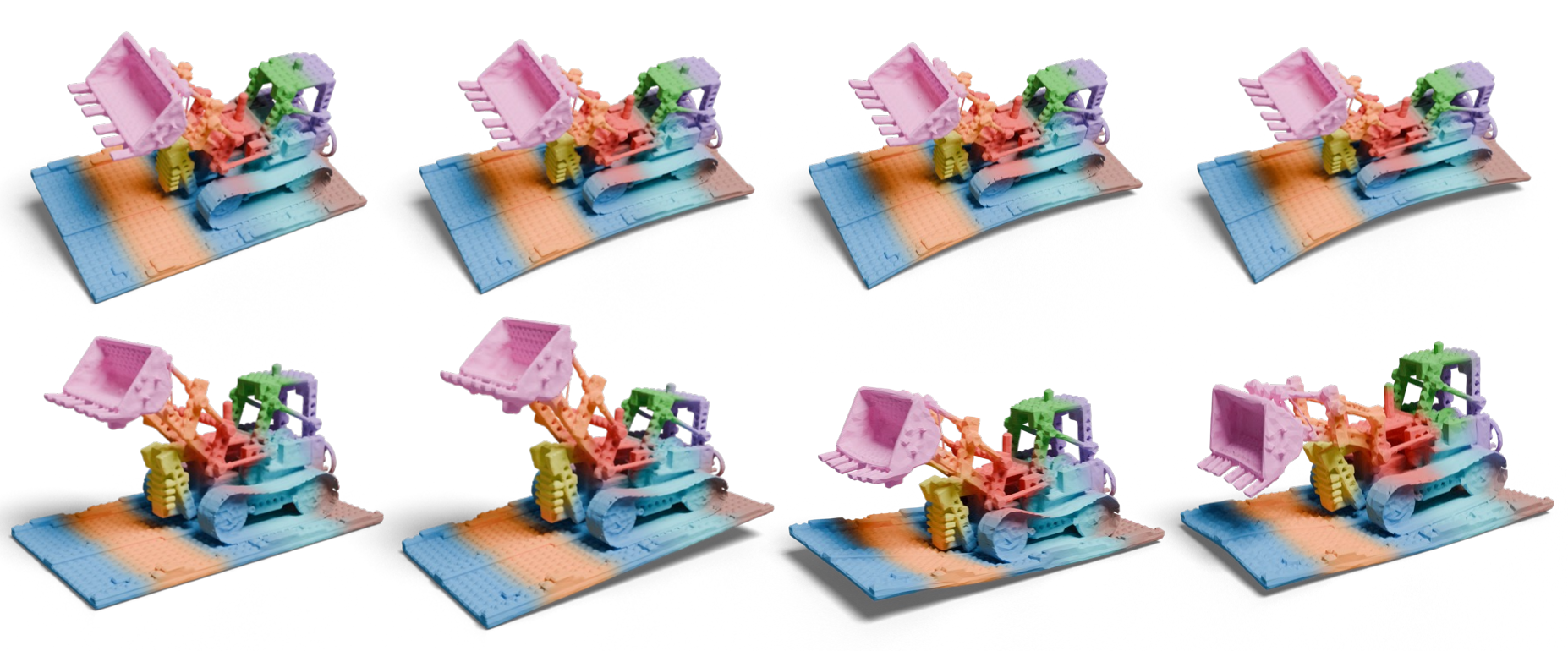}
    \vspace{0.1em}
    \caption{Skinning results for the deformable Lego shape family, where the deformable geometries are from Mani-GS~\cite{gao2024mani}.}
    \vspace{-0.5em}
    \label{fig:lego-family}
\end{figure}

\noindent\textbf{Deformable Shape Family.} In Fig.~\ref{fig:lego-family}, our method uses a single model to infer skinning fields for a deformable shape family, while Simplicits~\cite{modi2024simplicits} must be trained separately for each geometry. Our framework greatly improves efficiency and generalization across shapes.

\begin{figure}[t]
    \centering
    \vspace{-1.0em}
    \includegraphics[width=0.93\linewidth]{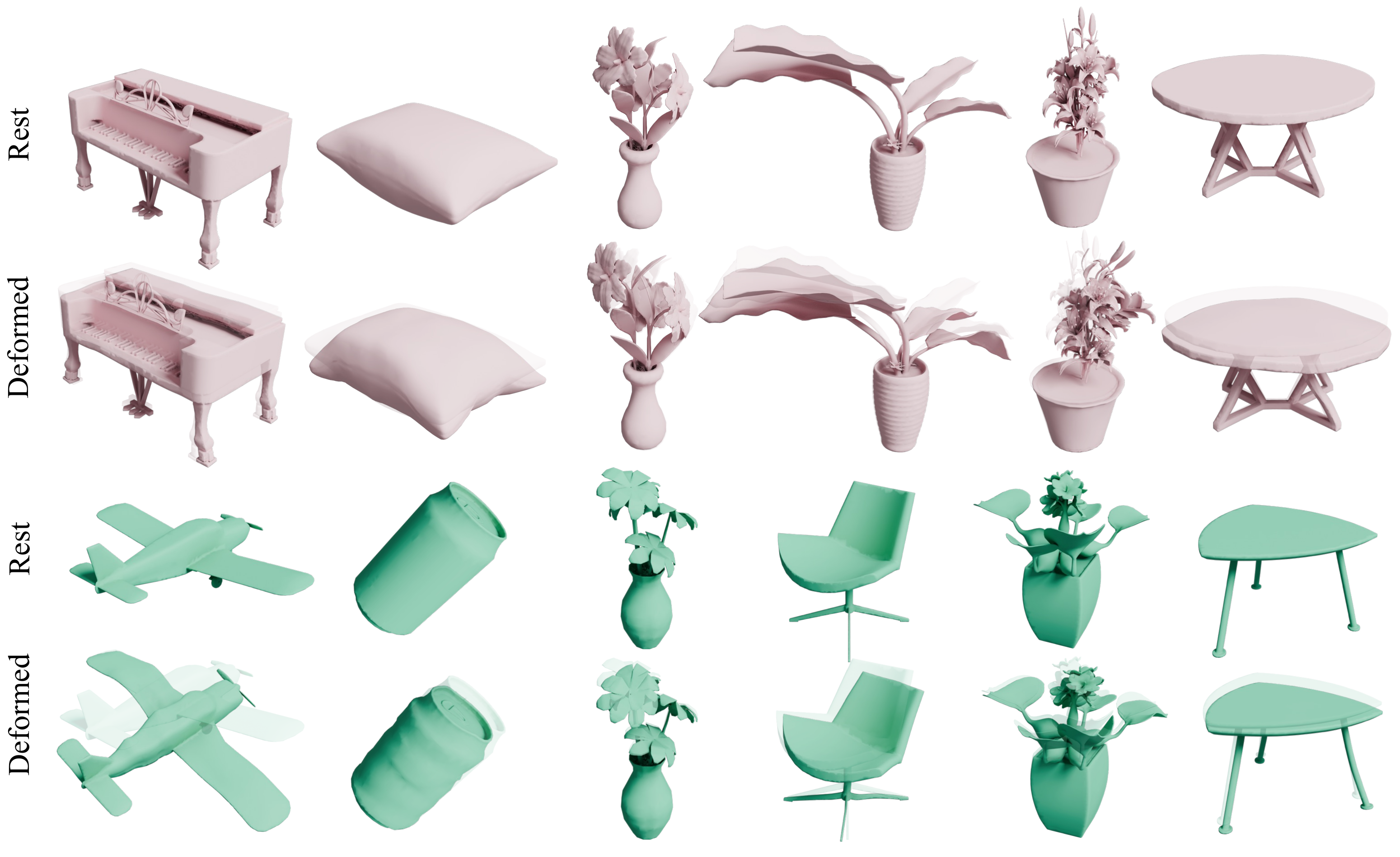}
    \caption{Animation results on ShapeNet~\cite{chang2015shapenet} mesh objects.}
    \vspace{0.8em}
    \label{fig:animation-results}
\end{figure}

\begin{figure}[t]
    \centering
    \vspace{-0.5em}
    \includegraphics[width=0.95\linewidth]{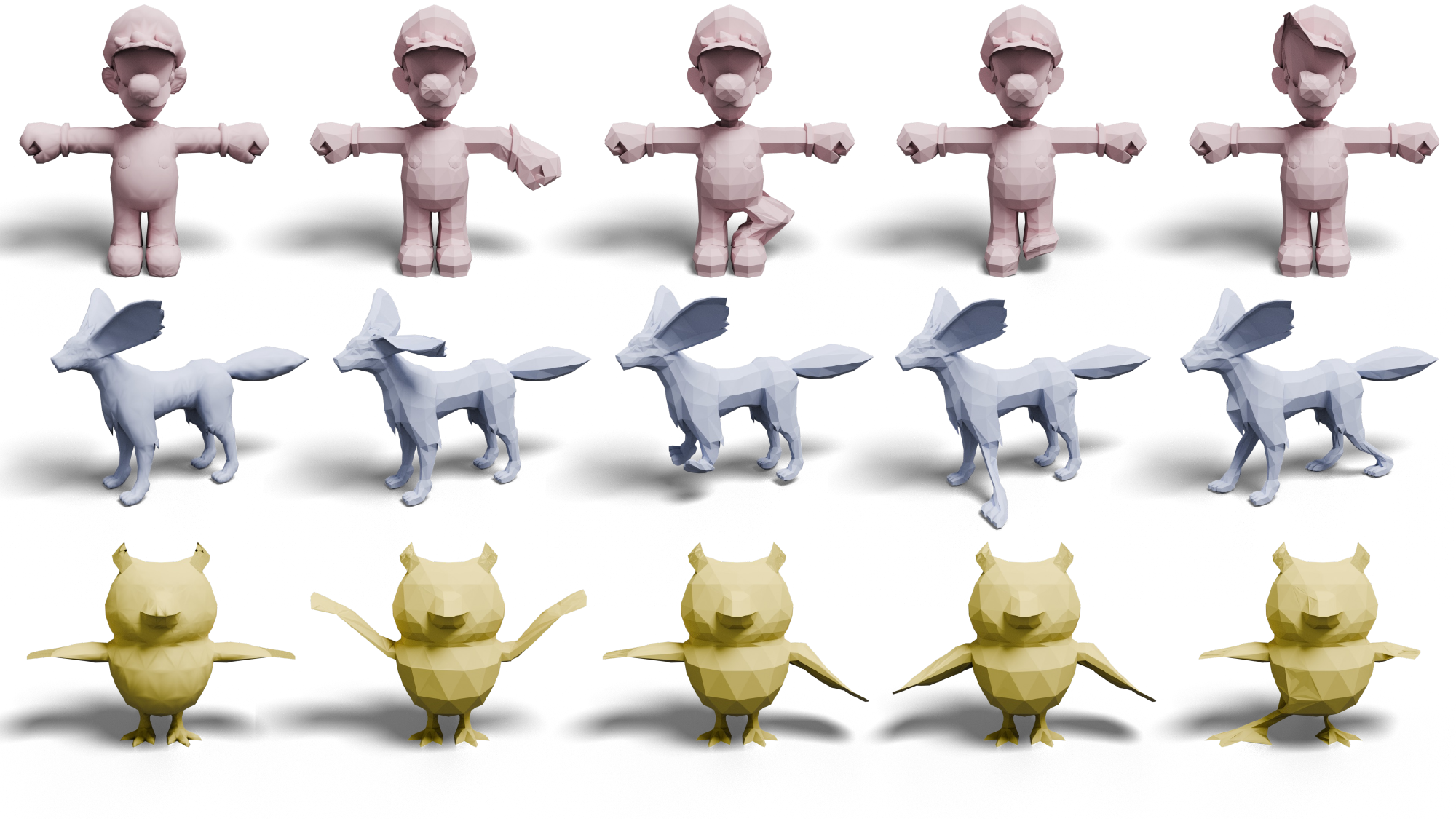}
    \caption{Animation results on RigNet~\cite{xu2020rignet} mesh objects.}
    \vspace{0.8em}
    \label{fig:animation-regnet}
\end{figure}

\begin{figure}[t]
    \centering
    \vspace{-0.5em}
    \includegraphics[width=0.92\linewidth]{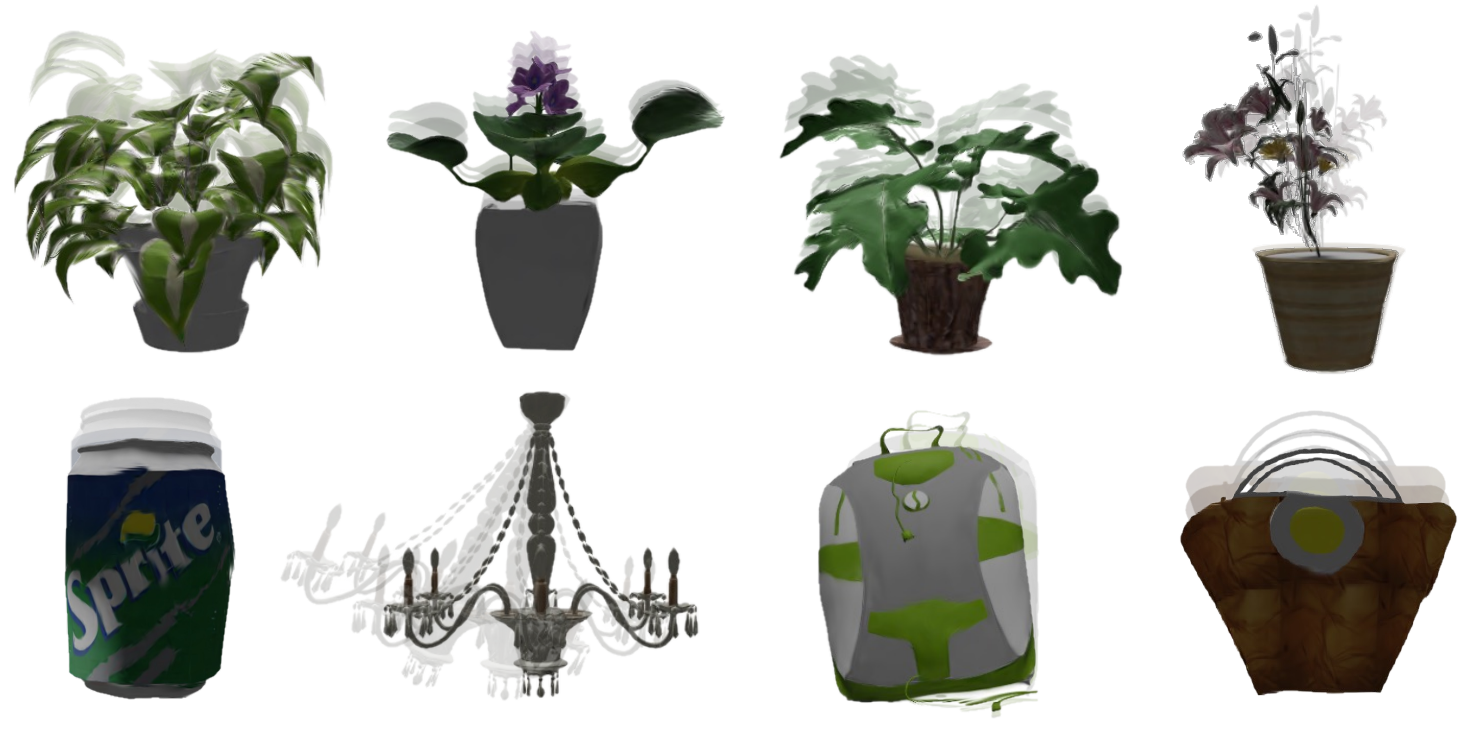}
     \caption{Animation results on 3DGS models~\cite{ma2024shapesplat}.}
    \vspace{-0.5em}
    \label{fig:animation-3dgs}
\end{figure}

\vspace{-0.2em}
\subsection{Application of Skinning-Based Animation}
\vspace{-0.2em}
\noindent\textbf{Physics-Based Animation.}  Our learned skinning fields can be seamlessly integrated with physics-based animation, as detailed in Sec.~\ref{subsec:skinning_eigenmodes}. In Fig.~\ref{fig:animation-results} and Fig.~\ref{fig:animation-regnet}, we present several physics-based animation results of our method on various 3D mesh models from the ShapeNet~\cite{chang2015shapenet} and RigNet~\cite{xu2020rignet} datasets. Additionally, owing to the discretization-agnostic property of our method, it can seamlessly animate static 3DGS models~\cite{ma2024shapesplat}, as demonstrated in Fig.~\ref{fig:animation-3dgs}. Our method can generate high-quality physics-based animations with complex deformations show that the learned skinning fields effectively capture the intrinsic deformation modes of the objects, enabling realistic and physically plausible animations under various external boundary conditions. For more animation results, please refer to the video supplementary.

\noindent\textbf{Time Efficiency.} We compare the time efficiency of our skinning-based subspace animation with classical implicit time integrator-based FEM method and explicit time integrator-based MPM method. For these two full-space physical simulators, we use open-source high-performance GPU implementations~\cite{chang2023licrom, zong2023neural}. All methods are evaluated under the same CPU and GPU hardware, as shown in Table~\ref{tab:eval_simtime}, our method demonstrates excellent time efficiency.

\begin{table}[t!]
\centering
\vspace{-0.2em}
\caption{Comparison of per-step simulation cost for physics-based animation with full-space simulators FEM~\cite{chang2023licrom} and MPM~\cite{zong2023neural}.}
\resizebox{0.9\linewidth}{!}{
  \setlength{\tabcolsep}{5pt}
  \footnotesize
  \begin{tabular}{lcccc}
    \toprule
    3D Shape &  \shortstack{Vertex\\ Count} & \shortstack{FEM~\cite{chang2023licrom}\\Step Cost (ms)} & \shortstack{MPM~\cite{zong2023neural}\\Step Cost (ms)} & \shortstack{Ours\\Step Cost (ms)} \\
    \midrule
    Airplane & 10K  & 79.83  & 141.83  & \textbf{12.26}   \\
    Bag & 121K & 3012.47  & 233.79   & \textbf{13.39}  \\
    Camera & 80K & 2121.02  & 203.38  & \textbf{12.52}  \\
    Chair & 52K & 1772.70 & 178.46  & \textbf{12.48} \\
    Pillow & 127K & 3170.93  & 251.81  & \textbf{13.74}  \\
    \bottomrule
  \end{tabular}
}
\vspace{0.5em}
\label{tab:eval_simtime}
\end{table}
\begin{table}[t!]
\centering
\caption{
Ablation studies on the RigNet~\cite{xu2020rignet} dataset.
}
\resizebox{0.95\linewidth}{!}{
\tabcolsep 8pt
\footnotesize
\begin{tabular}{lccc}
\toprule
\multicolumn{1}{c}{Config.} & \multicolumn{1}{l}{$\Omega_{\text{orth}} \times 10^{-2}$ $\downarrow$ } & \multicolumn{1}{l}{$\kappa_{\log}$ $\downarrow$} & \multicolumn{1}{l}{$H_{\text{spec}}$ $\uparrow$}  \\ 
\midrule
$\mathit{w/\!o}$ Skinning Normalization  & 6.5533  & 8.5492  & 0.8113  \\
$\mathit{w/\!o}$ ONI Layer & 0.0081 & 1.0844 & 0.9997     \\
$\mathit{w/\!o}$ ConFIG Optimization & 8.9247 & 11.8595  & 0.7594  \\
$\mathit{w/\!o}$ $\mathcal{L}_{\text{orth}}$ & 100.0  & 29.18  & NaN  \\
$\mathit{w/\!o}$ $\mathcal{L}_{\text{smooth}}$  & 0.0050  & 1.0567  &  0.9998  \\
Full Model & \textbf{0.0033} &  \textbf{1.0453} &  \textbf{0.9999}  \\
\bottomrule
\end{tabular}
}
\vspace{-0.9em}
\label{tab:ablations}
\end{table}

\begin{figure}[t]
    \centering
    \vspace{0.4em}
    \includegraphics[width=0.95\linewidth]{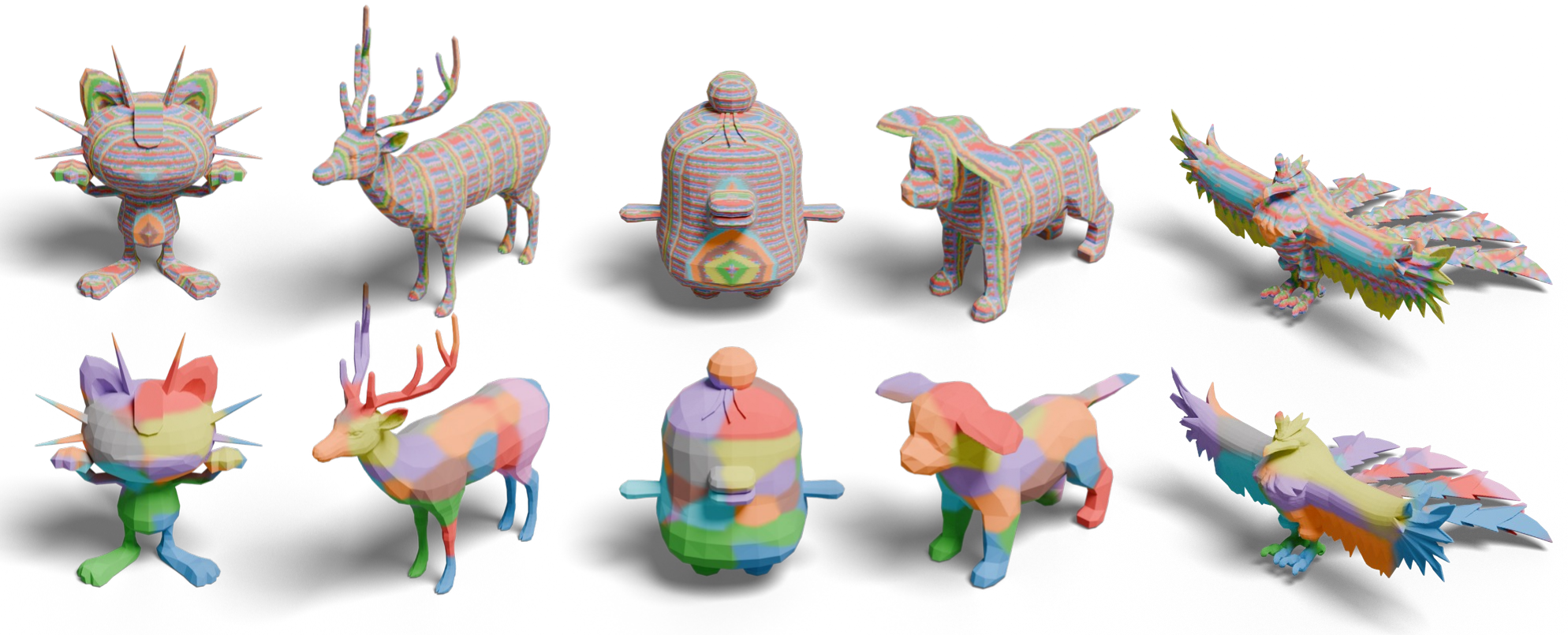}
    \caption{Top: Trained without $\mathcal{L}_{\text{pot}}$. Bottom: Trained with $\mathcal{L}_{\text{pot}}$.}
    \label{fig:ablation_energy}
\end{figure}

\vspace{-0.2em}
\subsection{Ablation Studies}
\vspace{-0.2em}
To validate our design choices, we conduct ablation studies on key components from the proposed pipeline and analyze their impact on performance using the RigNet~\cite{xu2020rignet} dataset. As shown in Tab.~\ref{tab:ablations}, we remove or modify key modules to assess their individual impact on model performance. Specifically, we evaluate the effects of the smoothness loss $\mathcal{L}_{\text{smooth}}$, the orthogonality constraint $\mathcal{L}_{\text{orth}}$, the skinning normalization process, the ONI~\cite{huang2020controllable} layer, and the conflict-free training (ConFIG)~\cite{liu2025config} stage. The quantitative results demonstrate that each component contributes to maintaining numerical stability and preserving the orthogonality of the learned neural skinning fields, which is essential for generating well-conditioned and physically consistent deformation behavior. In Fig.~\ref{fig:ablation_energy}, we visualize the skinning results obtained without the potential energy loss $\mathcal{L}_{\text{pot}}$, where the model fails to produce physically interpretable skinning fields.

\vspace{-0.8em}
\section{Conclusion}
\vspace{-0.6em}
In this paper, we present PhysSkin, a generalizable physics-informed  framework for real-time physics-based animation across diverse 3D shapes and discretizations. PhysSkin employs a new neural skinning fields autoencoder trained with a novel Physics-Informed Self-Supervised Learning (PISSL) strategy designed to enforce physical plausibility while ensuring numerical stability.
Experiments demonstrate that PhysSkin achieves outstanding performance on generalizable neural skinning, while delivering real-time performance on physics-based animation.
As a limitation, our method does not incorporate semantic priors for modeling complex geometries. 
Future work will incorporate such priors to further enhance the expressiveness of the model.

\noindent\textbf{Acknowledgment:} This work was supported by the NSFC No.~62572425, No.~624B2132, Information Technology Center, and State Key Lab of CAD\&CG, Zhejiang University.
\vspace{-1.5em}
{\small
\bibliographystyle{ieee_fullname}
\bibliography{egbib}
}

\end{document}